\documentclass[letterpaper, 10 pt, conference]{ieeeconf} 

\IEEEoverridecommandlockouts                              
\usepackage{graphicx} 
\usepackage{times} 
\usepackage{amsmath,bm} 
\usepackage{amssymb}   
\usepackage[hidelinks]{hyperref} 

\title{\LARGE \bf
Learning Agent Interactions from Density Evolution \\ in 3D Regions With Obstacles
}

\author{Amoolya Tirumalai, Christos N. Mavridis, John S. Baras
	\thanks{Department of 
	Electrical and Computer Engineering and the
	Institute for Systems Research, 
	University of Maryland, College Park, MD 20742, USA. 
	Emails: \{ast256, mavridis, baras\}@umd.edu}%
\thanks{This material is based upon work supported partially by ONR grant N00014-17-1-2622 (AT, CM, and JB), and by the Ann G. Wylie Dissertation Fellowship from the University of Maryland (Tirumalai).}
}

\begin{document}
\maketitle
\thispagestyle{empty}
\pagestyle{empty}

\begin{abstract}
In this work, we study the inverse problem of identifying complex flocking dynamics in a domain cluttered with obstacles. We get inspiration from animal flocks moving in complex ways with capabilities far beyond what current robots can do. Owing to the difficulty of observing and recovering the trajectories of the agents, we focus on the dynamics of their probability densities, which are governed by partial differential equations (PDEs), namely compressible Euler equations subject to non-local forces. We formulate the inverse problem of learning interactions as a PDE-constrained optimization problem of minimizing the squared Hellinger distance between the histogram of the flock and the distribution associated to our PDEs. The numerical methods used to efficiently solve the PDE-constrained optimization problem are described. Realistic flocking data are simulated using the Boids model of flocking agents, which differs in nature from the reconstruction models used in our PDEs. Our analysis and simulated experiments show that the behavior of cohesive flocks can be recovered accurately with approximate PDE solutions.
\end{abstract}

\section{Introduction}

In multi-agent (MA) systems, decisions depend critically on the interactions between agents (both local and global), the information each agent has on the dynamics and actions of other agents, and the topologies of various networks (or multigraphs) used to abstractly model the MA system (collaboration, information, and communication network) \cite{baras2014netscience}. Learning emerges naturally when addressing inference problems in MA systems associated with learning the dynamics of other agents \cite{mavridis2022learning, matei2019}, learning the interactions or coordination laws \cite{matei2019,mavridis2020learning,mavridis2020detection}, and learning to collaborate in decision making, inference or attention \cite{hovar2011, jiang2018}. 

A still-open problem in MA systems is how one can discover or characterize the interaction (or coordination) schemes that govern agent dynamics from data \cite{mavridis2020learning, mavridis2022learning, yang2022generative, hepworth2022swarm,mao2019nonlocal}. This problem is especially interesting in discovering how animals interact in flocks, swarms, and herds. 
Animals which move in these large collectives behave in emergent ways that are far more complex and useful than any behaviors achieved thus far by robots. 
They are also able to perform these maneuvers in very dense formations and within regions cluttered with obstacles. 
Beyond the interesting conclusions related to the basic science which one gains by discovering the interaction rules of such animals, one can also try to map these behaviors onto robots, especially when structured models are used. There are even more challenging and important problems where such inspiration and knowledge can be used, including social networks over the Internet, collaborating human-machine MA systems, and defence mechanisms
against aerial UAV swarm attacks \cite{mavridis2020detection}.

There are generally two broad classes of model-based approaches used to solve this problem: microscopic agent models, which have dynamics given by (typically nonlinear) ordinary or stochastic differential equations, and macroscopic models, which are described with partial differential equations
\cite{bongard2007automated, reynolds2022stochastic,mavridis2022learning}. 
However, one of the main problems in studying natural flocking lies in extracting information from data. Extracting useful and accurate trajectories of the agents has been shown to be very difficult \cite{mavridis2022learning,ballerini2008interaction}. 
%
On the other hand, extracting the empirical distribution of the dynamics (or more precisely, functionals of the empirical distributions, such as histograms) can be less prone to noise, as one extracts this information using operators which exhibit averaging properties. 
%
Hence, we focus on the macroscopic approach.
The macroscopic models are generally understood as mean-field limits (which can be in either a rigorous or formal sense) of microscopic model dynamics. The area of study which describes passage from microscopic to mesoscopic and then macroscopic models is described in kinetic and hydrodynamic theory, and is a thoroughly studied area of research in applied mathematics \cite{carrillo2010particle,mao2019nonlocal}.

\subsection{Contribution.}
Our objective in this work is to discover the interaction (or coordination) dynamics of a large swarm of agents interacting within a workspace cluttered with obstacles, 
via observations of their density, i.e., their probability distribution in three spatial dimensions, plus time.

We generate 3D particle trajectories using a version of the Boids model \cite{reynolds1987flocks}. Such models have been used extensively to generate very realistic scenes and system trajectories and associated data. We bound the Boids to a cubical region of space with obstacles formed from cubes, and subject them to specular reflections. 
From these trajectories, we extract the histograms of the density and momentum profiles. 
We represent the initial mean-field distribution and momentum profile as a deep neural network for the initialization of the PDE describing the mean-field spatial probability distribution.
The time-evolving spatial probability distribution estimates are then used
as continuous observations of the density of the swarm.
To discover the interaction (or coordination) laws, which we parametrize with a real vector, we pose a PDE-constrained optimization problem to minimize the squared Hellinger distance between the Boid histogram and the hydrodynamic mean-field distribution. We solve this problem numerically and approximately using the Newton-conjugate gradient method. 
To the authors' knowledge, this is the first instance of a macroscopic 
MA system identification problem for flocking being solved in three spatial dimensions
in the presence of obstacles. 
%
%

\subsection{Related Work}

In our previous work, we discussed the tradeoffs associated to particle and density-based approaches to discovering swarm interaction laws \cite{mavridis2020learning}, and explored this problem in two spatial dimensions, plus time \cite{mavridis2022learning}. 
Accurate and efficient solvers for the hydrodynamic Cucker-Smale model were developed which employed the Kurganov-Tadmor finite volume method to compute the hyperbolic part of the PDEs, and spectral solvers which computed the nonlocal part. 
In \cite{tirumalai2021weak}, we rigorously studied the hydrodynamic model that will be adopted and modified slightly in this work. We established existence and weak-strong uniqueness for this model in smooth regions. 
Our work was influenced by \cite{mao2019nonlocal}, where the authors followed a similar approach modeling the interactions with respect to a fractional differential system of equations . 
More recently, the same problem is studied from the perspective of physics-informed neural networks \cite{yang2022generative}. 

\subsection{Notation}

Throughout this work, we use `$...$' to indicate line continuation for long equations or expressions. 
We use the following notations for the inner product and outer products for real vectors $v,w \in \mathbb R^3$:
\[
v^\top w \equiv \langle v,w\rangle_{\mathbb R^3}, 
vw^\top \equiv v \otimes w,
\]
where $v^\top$ is the transposition of $v$. 
We use $l:\mathbf L(S)\rightarrow \mathbb R^+_0$ to denote the standard Lebesgue measure, $\mathbf L(S)$ being the $\sigma-$algebra of Lebesgue-measureable sets contained in $S$. If $S$ is bounded and smooth(-enough), $\mathbf n: \partial S \rightarrow \partial B(0,1)$ is the outward pointing normal on the boundary $\partial S$. We use $\mathbb I_{A}:\mathbb R^3\rightarrow \{0,1 \}$ to denote the characteristic (indicator) function of a set $A$. For some countable set $B$, $|B|$ is the cardinality of that set. Given a vector of $N$ position-velocities $(\mathbf x, \mathbf v) \in \mathbb R^{6N}$, the neighborhood of a point $x^i \in \mathbb R^3$ is:
\[
N^i(\epsilon):= \{x^j : ||x^i - x^j||_{\mathbb R^3} \leq \epsilon\}.
\]
For a set $S$, $\mathcal B(S)$ is the Borel $\sigma-$algebra on $S$.

\section{Model Formulation}
In this section, we describe the physical area where we assume agents are free to move, the model we use to generate synthetic data, and the model which we use to perform system identification.

\subsection{Workspace}

We assume that all agents are contained in a bounded polyhedral region $D \subset \mathbb R^3$. We assume that we have:
\[
D_m:= \{x \in \mathbb R^3 : ||x-\xi_m||_{\infty} \leq R_m\}, \xi_1 = 0
\]
and
$
 D_m \cap D_n = \emptyset, 2 \leq m,n \leq M,
$
$R_1 > R_m, 2 \leq m \leq M$, $\bigcup_{m=2}^M  D_m \subset D_1$. Then, we define:
\[
D:=  D_1 \backslash \bigcup_{m=2}^M  D_m.
\]

\subsection{Boids}

We generate flocking data according to a version of the Boids model \cite{reynolds1987flocks}. Consider agent position-velocity pairs \[(\mathbf x, \mathbf v):[0,T] \rightarrow  D^N \times \mathbb R^{3N}, (x^i, v^i): [0,T] \rightarrow  D \times \mathbb R^3,\] subject to ODEs:
\begin{equation}
	\begin{split}
	\label{boids}
	&\frac{d}{dt}x^i(t) = v^i(t); \\
	&\frac{d}{dt}v^i(t) = \sum_{j=1}^5 \mathbf F_j^i(\mathbf x(t), \mathbf v(t)); \\
	& (x_0^i, v^i_0) \sim \mathbb P(0),
	\end{split}
\end{equation}
where $\mathbb P(0) \in \mathcal P(D)$, is a probability measure on $D$ a.c. w.r.t. the Lebesgue measure. The forces are:
\[
\mathbf F_1^i(\mathbf x, \mathbf v):= \frac{1}{|N^i(1)|}\sum_{k=1}^{|N^i(1)|}(v^k - v^i);
\]
\[
\mathbf F_2^i(\mathbf x, \mathbf v):= \frac{1}{|N^i(2)|}\sum_{k=1}^{|N^i(2)|}(x^k - x^i);
\]
\[
\mathbf F_3^i(\mathbf x, \mathbf v):= \frac{1}{|N^i(\frac{1}{2})|}\sum_{k=1}^{|N^i(\frac{1}{2})|}\frac{x^i - x^i}{||x^i - x^k||_{\mathbb R^3}^2};
\]
\[
\mathbf F_4^i(\mathbf x, \mathbf v):= - (\frac{1}{16}||v^i||^2_{\mathbb R^3} - 1)v^i;
\]
\[
\mathbf F_5^i(\mathbf x, \mathbf v):= - 4\sum_{k=1}^M \frac{\mathbb I_{r < 1}(d(x^i, \partial D_m))}{d(x^i, \partial D_m)}\frac{x^i - \xi_k}{||x^i - \xi_k||_{\mathbb R^3}}.
\]
It is also assumed that each particle is subject to specular reflections:
\[
	\lim_{t^{*+}_{i,j} \downarrow t_{i,j}^*} v^j(t^{*+}_{i,j})=  
	(\mathbf I_3 - 2\mathbf n(x^j(t^{*}_{i,j}))\mathbf n^\top(x^j(t^{*}_{i,j}))) v^j(t^{*}_{i,j}),  
\]
where $t^{*}_{i,j} \in (0,T]$ is the time of the $i$-th collision of the $j$-th agent with the boundary $\partial D$. 
\subsection{Hydrodynamic Model}
In the interest of saving space, we direct the reader to our previous work on the Euler Alignment System for a description of the process one takes to obtain, at least in form, a hydrodynamic model with non-penetrative boundary conditions from a particle-based model with specular reflections \cite{tirumalai2021weak}. This approach is based on \cite{bae2019flocking, desvillettes2005trend}. The probability density of a collective of particles $\rho:[0,T] \times D \rightarrow \mathbb R$ and the momentum density $\mathbf j: [0,T] \times D \rightarrow \mathbb R^3$ evolve according to non-locally forced compressible Euler equations:
\begin{equation}
	\begin{split}\label{euler2}
		&\partial_t \rho + \nabla_x \cdot \mathbf j = 0 \text{ in } (t_0,t_f] \times D; \\
		&\partial_t \mathbf j + \nabla_x \cdot [\rho^{-1}\mathbf j \mathbf j^\top] = 
		\mathbf S[\rho, \mathbf j] \text{ in } (t_0,t_f] \times D; \\
		&\Lambda_a \mathbf \pi_\rho[\rho](t,\cdot) = \mathbb I_D \rho(t,\cdot) \text{ in } \mathbb R^3;\\
		&\Lambda_a \mathbf \pi_{\mathbf j}[\mathbf j](t,\cdot)= \mathbb I_D \mathbf j(t,\cdot) \text{ in } \mathbb R^3; \\
		&\Lambda_c v_c[\rho](t,\cdot) = \mathbb I_D \rho(t,\cdot) \text{ in } \mathbb R^3;  \\
		&\Lambda_r v_r[\rho](t,\cdot) = \mathbb I_D \rho(t,\cdot) \text{ in } \mathbb R^3; \\
		&\mathbf S[\rho, \mathbf j] := \rho \pi_{\mathbf j}[\mathbf j] - \mathbf j \pi_\rho[\rho] - ... \\ 
		&\rho \nabla_x[\mathbf 1^\top_2 \mathbf V[\rho] + U] 
		+ 
		k_p \mathbf j (1-F(\rho^{-1}||\mathbf j||_{\mathbb R^3} )); \\
		&\mathbf j^\top \mathbf n = 0 \text{ in } [t_0,t_f] \times \partial D; \\ 
		&(\rho(t_0,\cdot), \mathbf j(t_0,\cdot))^\top = (\rho_0, \mathbf j_0)^\top \text{ in } D,
	\end{split}
\end{equation}
 where $\mathbf V[\rho]:=(v_c[\rho], v_r[\rho])^\top$, and the operators $\Lambda_{(\cdot)}$ are:
 \[
\frac{1}{4 \pi k_{(\cdot)}} (\nabla_x^2 - \lambda_{(\cdot)})
 \]
 which is a Bessel-type operator with fundamental solution \cite{tirumalai2021weak}:
 \[
 G(x,s;k_{(\cdot)},\lambda_{(\cdot)}) = \frac{k_{(\cdot)}e^{- \lambda_{(\cdot)}||x-s||_{\mathbb R^3}}}{||x-s||_{\mathbb R^3}}.
 \]
We do not specify concretely the domains and ranges of these operators, as we feel that the functional analysis required to define them clearly do not contribute much to the thrust of the paper, which is application-focused. Those who desire details are encouraged to refer to \cite{tirumalai2021weak}. For vector functions, take the operator to apply element-wise.

We take $F(s):= 1 + \tanh(\frac{s^2}{\lambda_p^2} - 1)$. Finally, we take:
\[
U(\cdot) := k_o\eta(\cdot;\omega_o)*(\mathbb I_{d(\xi,\partial D_1) < \omega_o}(\cdot) + \sum_{i=2}^M\mathbb I_{\xi \in D_i}(\cdot)),
\]
where $\eta(\cdot;\epsilon)$ is the standard mollifier/ bump function \cite{evans2010partial}. 
\section{Problem Formulation}

In this section, we frame the optimization problems used for system identification. First, let us clearly define some quantities. For two probability measures $P,Q \in \mathcal P(D)$ which are a.c. w.r.t. the Lebesgue measure on $D$, the squared Hellinger distance is:
\[
H^2(P,Q):= \frac{1}{2} \int_D (\sqrt{p(x)} - \sqrt{q(s)})^2 d l(x),
\]
where $p,q$ are the densities (Radon-Nikodym derivatives) of $P,Q$ w.r.t. the Lebesgue measure, respectively. Let $v_{max}:= \max_{(t,i) \in [0,T] \times \{1,...,N\}} ||v^i(t)||_\infty$, , and define a sequence of regions: $\{\Omega_{ijk} \times \Upsilon_{lmn}\}_{i,j,k,l,m,n=1}^\mathfrak N$ which cover $[-R_1,R_1]^3 \times [-v_{max},v_{max}]^3$ where $\Omega_{ijk}$ has volume $\Omega := (2R_1 \backslash \mathfrak N)^3$, and $\Upsilon_{lmn}$ has volume $\Upsilon:= (2 v_{max} \backslash \mathfrak N)^3$. Now, we define the position-velocity histogram over these cells as:
\[
\mu^N(t,\Omega_{ijk},\Upsilon_{lmn}):= \frac{1}{N} \sum_{\nu=1}^N \mathbb I_{(x,v) \in \Omega_{ijk} \times \Upsilon_{lmn} }(x^\nu(t), v^\nu(t)).
\]
Via the histogram, we define a position-velocity density as:
\begin{equation*}
	\begin{split}
&m^N(t,x,v):= ... \\ 
&\frac{1}{\Omega \Upsilon}\sum_{i,j,k,l,m,n=1}^\mathfrak N \mathbb I_{(\xi,\upsilon) \in \Omega_{ijk} \times \Upsilon_{lmn} }(x, v) M^N(t,\Omega_{ijk},\Upsilon_{lmn}),
\end{split}
\end{equation*}
associated to measure for $A \in \mathcal B([-R_1,R_1]^3 \times [-v_{max},v_{max}]^3)$:
$$
M^N(t,A):= \int_A m^N(t,x,v) dl(x,v).
$$
 The position histogram is:
\[
q(t,x):= \int_{v \in \mathbb R^3} dM^N(t,x,v)
\]
and the associated measure is:
\[
Q(t,A):= \int_A q(t,x) dl(x),
\]
with $A \in \mathcal B([-R_1,R_1]^3)$.
For a solution to the PDE (\ref{euler2}) at a particular time $t \in [0,T]$, define a measure (assuming $\rho(t,\cdot)$ is extended by $0$ from $D$ to $D_1$):
\begin{equation}
	\label{measure}
P(t,A):= \int_A \rho(t,x) dl(x)
\end{equation}
for $A \in \mathcal B([-R_1,R_1]^3)$.
\subsection{Initial Conditions}
For the initial conditions, we assume that they are of the form:
\[
(\rho_0,\mathbf j_0)^\top = (\rho_0(\cdot;\theta), \mathbf j_0(\cdot;\theta))^\top
\]
where $\theta \in \mathbb R^d$ is a vector of neural network weights, i.e. the initial conditions are members of a family of neural network-parametrized functions. The neural network structure selected is that of \cite{al2019extensions} called the ``Deep Galerkin Method'' (DGM), which is similar to the LSTM structure \cite{vanhoudt2020}. The input is 3D, there are 2 DGM hidden layers with 50 DGM neurons each, and an output layer with 4 outputs, 1 for the density, and 3 for the momentum. This results in $d = 20350$ weights. The hidden layer neurons were {\it tanh} neurons, and the outputs were a {\it softplus} neuron for the density, and {\it tanh} neurons for the momentum.  Let $P_0(\cdot;\theta)$ be the probability measure with $\rho_0(\cdot,\theta)$ as its density. The learning problem we solve for the initial conditions is:
\begin{equation}
	\begin{split}
\min_{\theta \in \mathbb R^d}& L_1(\theta) := H^2(P_0(\cdot;\theta), Q(0,\cdot)) + ||\mathbf j_0(\cdot;\theta) - \hat{\mathbf j}_0||_{\mathbb R^3}^2 \\
&\text{s.t. } \int_D dP_0(x;\theta) = 1, \rho_0(\cdot;\theta) \geq 0
\end{split}
\label{eq:nn-loss}
\end{equation}
where 
\[
\hat{\mathbf j}_0:= \int_{v \in \mathbb R^3} v dM^N(0,x,v).
\]
This deep learning problem is solved in the typical way using the stochastic gradient method ADAM \cite{kingma2014}. We scheduled the learning rate inversely w.r.t. the iteration number. The DGM network was structured so as to satisfy the constraints specified.
\subsection{System Identification}
In the model (\ref{euler2}), there are $\hat d=10$ parameters, which we denote by $\theta:= (k_a, k_c, k_r, k_p, k_o, \lambda_a, \lambda_c, \lambda_r, \lambda_p, \lambda_o)$. We pose the following PDE-constrained optimization problem:
\begin{equation}
	\begin{split}
		\min_{\theta \in \mathbb R^{\hat d}}& L_2(\theta):=\frac{1}{t_f - t_0}\int_{t_0}^{t_f}H^2(P(t,\cdot;\theta), Q(t,\cdot)) dt \\ 
		&\text{s.t. } (\ref{euler2}), (\ref{measure}).
	\end{split}
 \label{eq:pde-loss}
\end{equation}

We solve this problem numerically using a Newton-conjugate gradient method as we did in \cite{mavridis2020detection}.
\section{Numerical Methods}

\subsection{Numerical Integration}

All integrals in this work are approximated simply using Riemann integrals. Take the discretization of $D$ to be given by $\{\Omega_{ijk}\}_{i,j,k=1}^\mathfrak N$, with centroids $\{s_{i,j,k}\}_{ijk=1}^N$. In (\ref{euler2}), notice the sequence of elliptic problems. Via the fundamental solution, for example, the first problem is solved as:
\[
\pi_\rho[\rho](t,\cdot) = \int_{\mathbb R^3} G(\cdot,s;k_a,\lambda_a) \rho(t,s)ds,
\]
which, due to the structure of $G(\cdot;k_a,\lambda_a)$, is a singular convolution integral. With a slight abuse of notation, denote:
\[
G(x,s;k_a, \lambda_a) = G(x-s;k_a,\lambda_a).
\]
In discretization, the convolution integral becomes the Riemann (convolution) sum:
\[
\pi_\rho[\rho](t,\cdot) = \sum_{i,j,k=1}^\mathfrak N \tilde G(\cdot-s_{i,j,k};k_a,\lambda_a)\rho(t,s_{i,j,k}) \Omega
\]
where (assuming $\lambda_a \neq 0$:
\[
\begin{split}
&\tilde G(x-s;k_a,\lambda_a) = ... 
\\& \begin{cases}
3 k_a\frac{1 - \exp(- \lambda_a \frac{R_1}{2 \mathfrak N})(\lambda_a \frac{R_1}{2 \mathfrak N}+1)}{\lambda_a^2 (\frac{R_1}{2 \mathfrak N})^2 } &\text{ if } ||x-s||_{\mathbb R^3} < \frac{R_1}{2 \mathfrak N}, \\
G(x-s;k_a,\lambda_a) &\text{ else }.
\end{cases}
\end{split}
\]
so as to deal with the singularity when $x=s$. The first branch is constructed by averaging the value of $G(\cdot;k_a,\lambda_a)$ over $[-R_1,R_1]$. To calculate the convolution sum efficiently, we take the very typical FFT-based approach.


\subsection{Numerical ODEs and Collision Event Handling}

To simulate the ODE system (\ref{boids}), we employ the velocity-Verlet \cite{mao2019nonlocal} method, commonly used in particle dynamics. To deal with the collisions of agents with the boundary of the region $D$, we employ a typical approach taken in gas particle dynamics and molecular dynamics. The regions we deal with are cuboidal, so identifying if a particle has penetrated them in a timestep is rather simple. Between timesteps, particles are assumed to travel along a ray. To correct a boundary penetration, we trace this ray back to when it intersects the boundary of the region $D$. Again, as $D$ is constructed from cubes, this is rather trivial. The time when this occurs is also calculated, and the particle is marched in time up to the collision. Then, the particle's velocity is reflected about the surface normal, easily determined as the boundaries of the region $D$ are entirely flat, and then the particle's position is marched up to the next timestep. Thus, the specular reflection condition specified is satisfied. Inter-agent collisions are not considered.

\subsection{Numerical PDEs}

In short, we employ a three-dimensional analogue of the second-order finite volume method we describe in our work in \cite{mavridis2022learning}. The 3D version of this method can be obtained by simply adding an extra index to the arrays described there. We do modify the fluxes in this model slightly. This is the Kurganov-Tadmor \cite{kurganov2000new} flux with the minmod flux-limiter modified with the positivity-preserving flux-limiter of \cite{zhang2012positivity}. The positivity preserving flux employed is the classical Rusanov-Local Lax-Friedrichs flux. In addition, we employ adaptive time-stepping so that the CFL number of the simulation is kept below $.05$. Time-marching is done via the strong stability-preserving second-order Runge-Kutta method \cite{kurganov2000new}.

There is also a small difference in the ``dissipation'' term $\mathbf S[\cdot]$ in this model. In our previous work, we only applied an alignment term. In this work, we deal with several more terms in the functional $\mathbf S[\cdot]$. The arrays which contain the discretizations of the terms of $\mathbf S[\cdot]$ are simply added together.

\subsection{Choice of Parameters}

For the workspace, we take $4$ cubic obstacles, each of size $R_2 = ... = R_5 = 1$. For the outer boundary, we let $R_1 = 5$. The centers of the obstacles were the columns of:
\[
\begin{bmatrix} 2.5 & 2.5 & -2.5 & -2.5 \\
				2.5 & -2.5 & 2.5 & -2.5 \\
				-4 & -4 & -4 & -4
	\end{bmatrix}.
\]
so each obstacle rests on the bottom of the arena. As stated earlier, timestepping is adaptive. The spatial mesh was taken to be $11 \times 11 \times 11$. We simulated 2000 Boids. Their initial conditions were sampled from a Gaussian, and re-sampled if they lay outside $D$.

\section{Results}

\begin{figure}
	\centering
	\includegraphics[scale=.42,trim={0 0 0 20}]{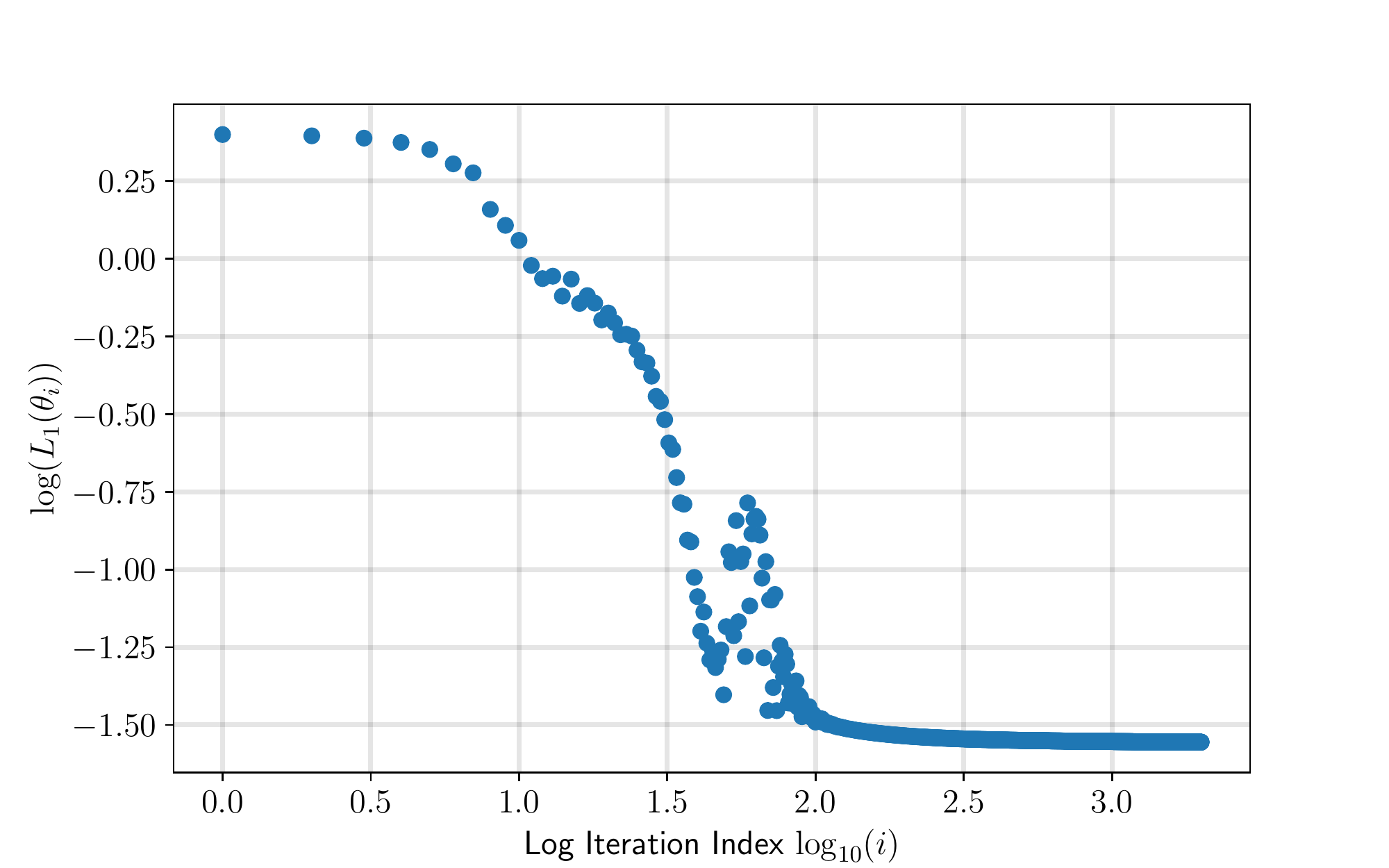}
	\caption{Training loss $L_1$ of the deep learning approximation problem \eqref{eq:nn-loss} representing the initial conditions of the distribution of the agents simulated with the Boids 
model. 
}
 \label{fig:nn-loss}
\end{figure}

\begin{figure}
	\centering
	\includegraphics[scale=.42,trim={0 0 0 20}]{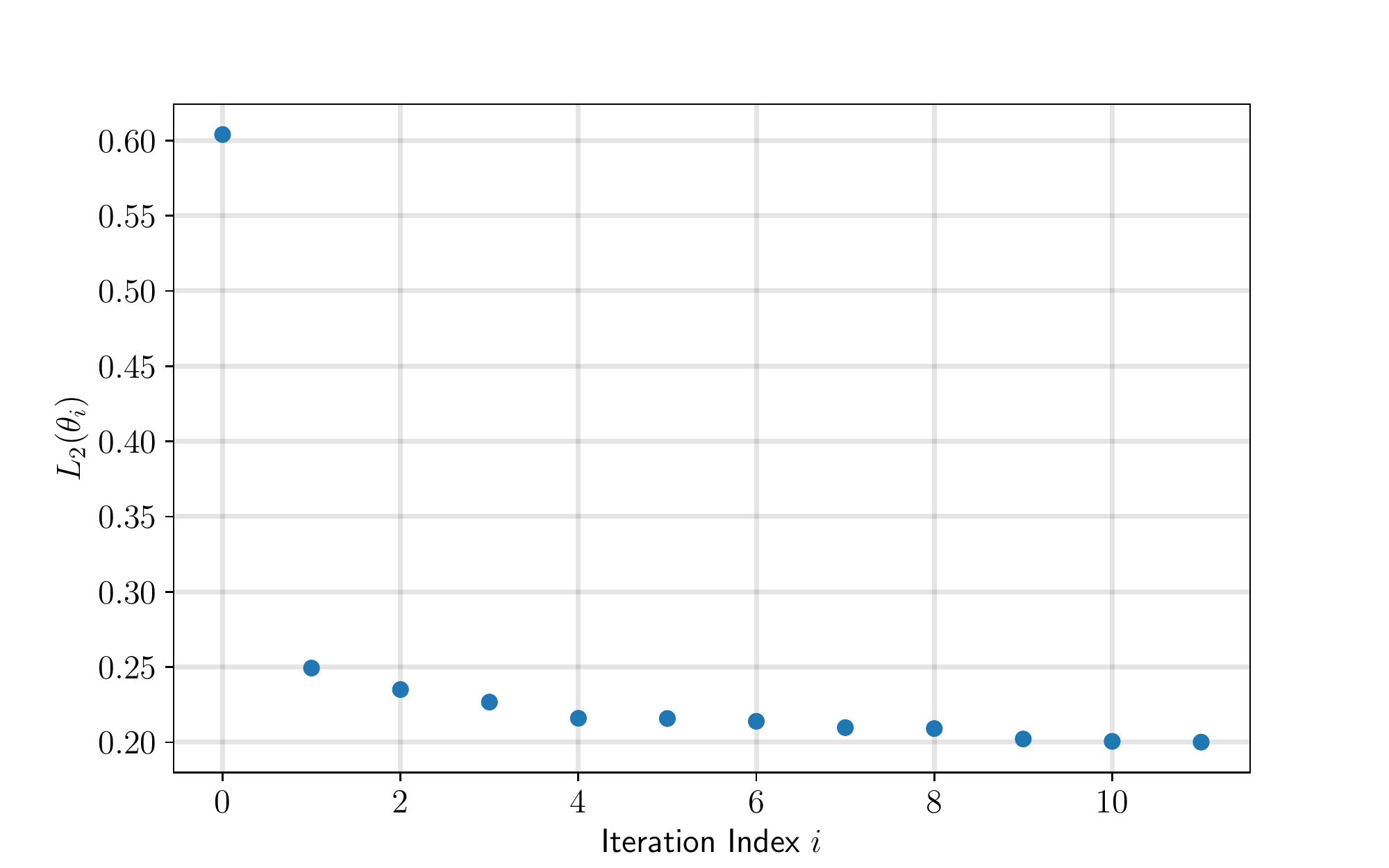}
	\caption{The loss functional in the PDE-constrained optimization problem
 \eqref{eq:pde-loss}. The initial descent steps of the Newton method are quite steep, in accordance to our observations in our previous work \cite{mavridis2020learning}.}
\label{fig:loss-pde}
\end{figure}

The initial condition of the distribution of the agents simulated with the Boids 
model are represented with a deep neural network. The training error in the optimization process is depicted in Fig. \ref{fig:nn-loss}. 
In Fig. \ref{fig:loss-pde}, the loss in our system identification problem is plotted against the iteration index. The Newton-conjugate gradient method brought quickly the PDE-constrained optimization problem we posed to a suboptimal solution. 
The Newton iterations terminated after $11$ iterations. 
The final loss was $L_2(\theta_{11}) = 0.20008555429515784$.

\begin{figure}
	\centering
	\includegraphics[scale=.42,trim={0 0 0 20}]{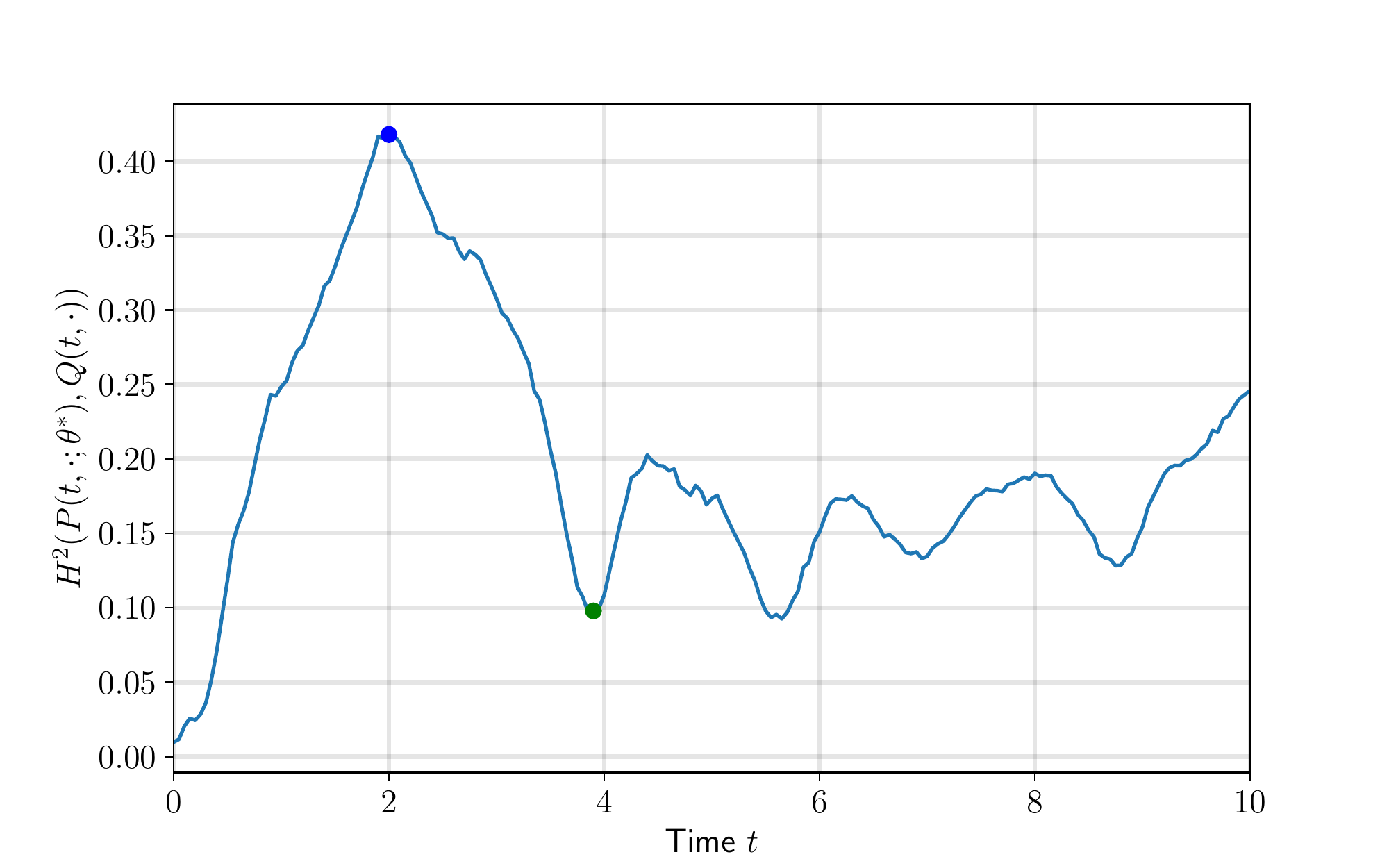}
	\caption{The squared Hellinger distance between the Boid distribution and the mean-field distribution at each simulation time. The mean-field distribution is interpolated linearly in time to match those of the Boid simulation as it is timestepped according to its own stability criteria. We have highlighted two points. In blue, we highlight the time (time instant $t\simeq 2$) at which the boid and mean-field distributions are the most different. In green, we highlight a time (time instant $t\simeq 4$) when they are more similar. We plot these distributions in Fig. \ref{fig:boids-2} and Fig. \ref{fig:boids-4}, respectively.}
 \label{fig:hellinger}
\end{figure}

\begin{figure}
	\centering
	\includegraphics[scale=.5]{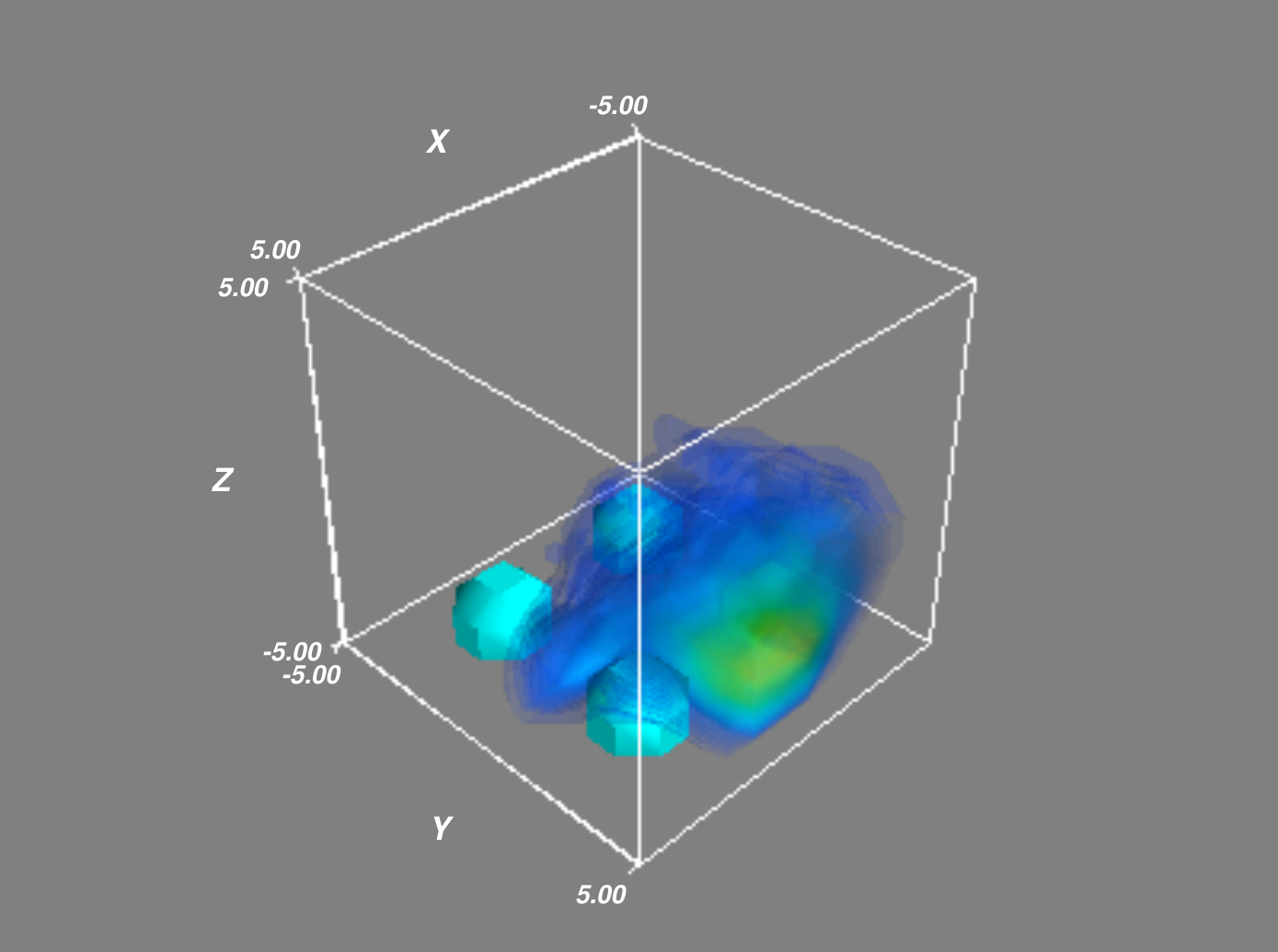}
 	\centering
	\includegraphics[scale=.5]{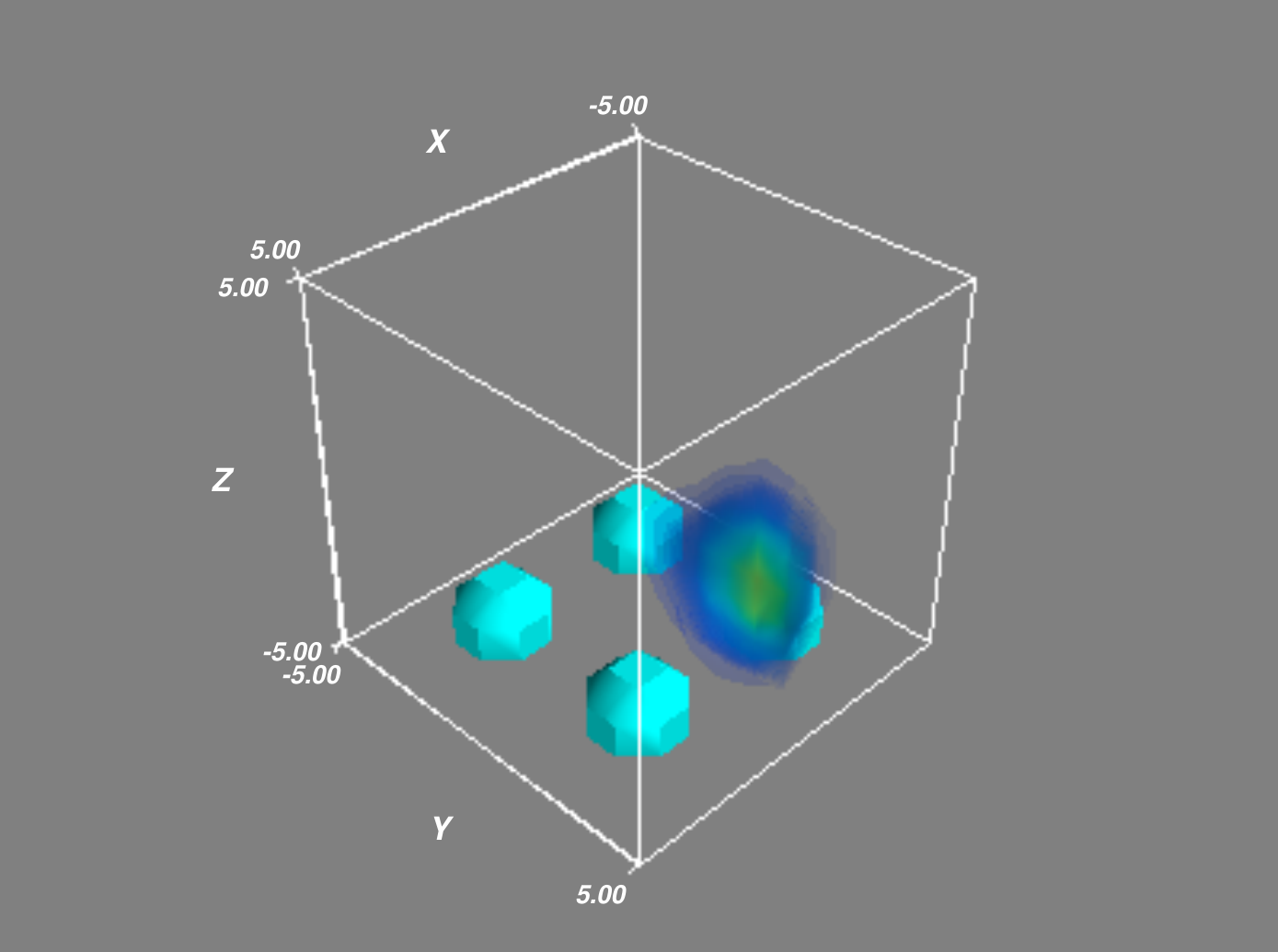}
	\caption{(top) The Boid distribution at the time (time instant $t\simeq 2$) of maximal Hellinger distance from the mean-field distribution. (bottom) The mean-field distribution at the same time.}
 \label{fig:boids-2}
\end{figure}

\begin{figure}
	\centering
	\includegraphics[scale=.5]{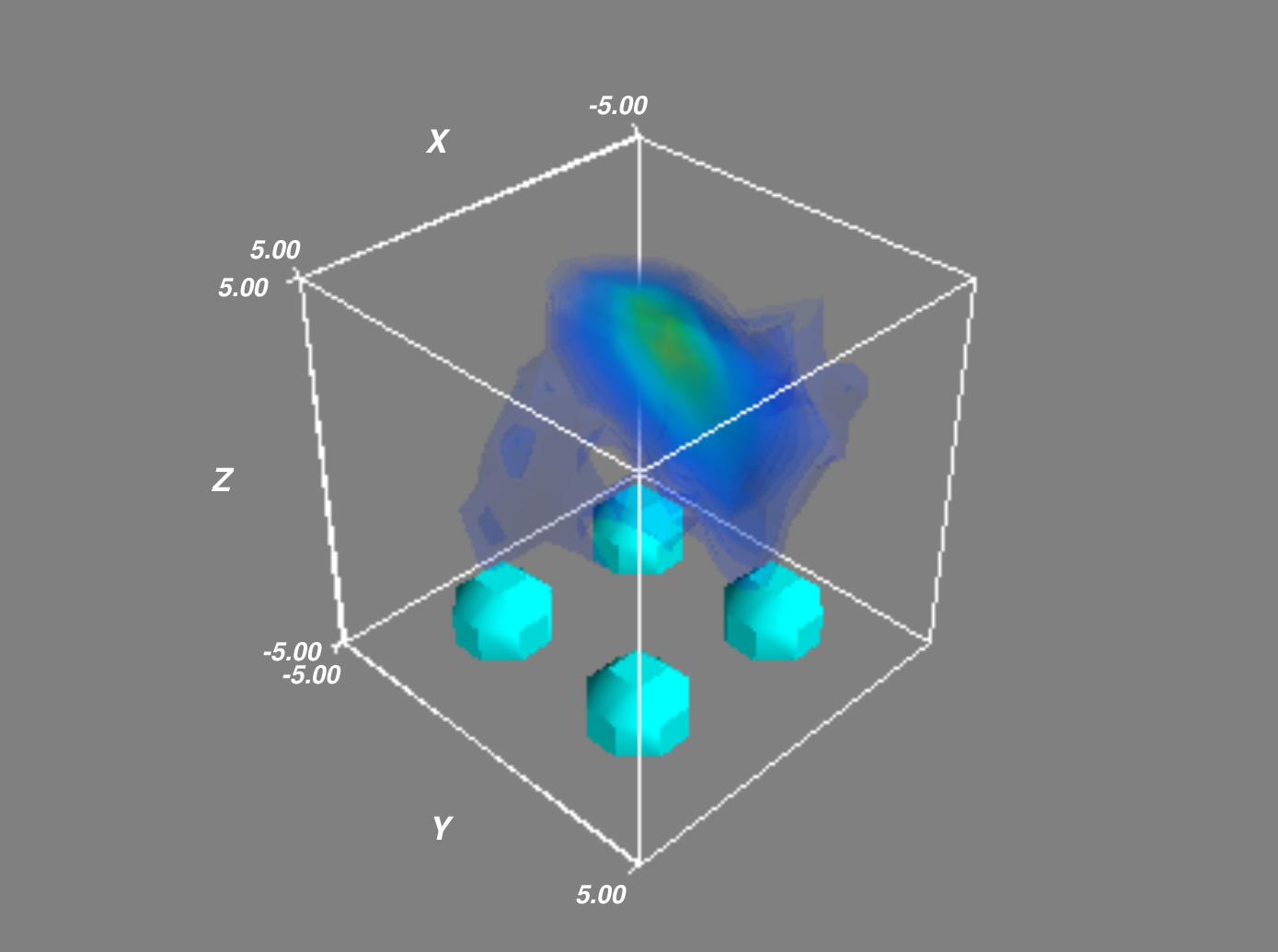}
 	\centering
	\includegraphics[scale=.5]{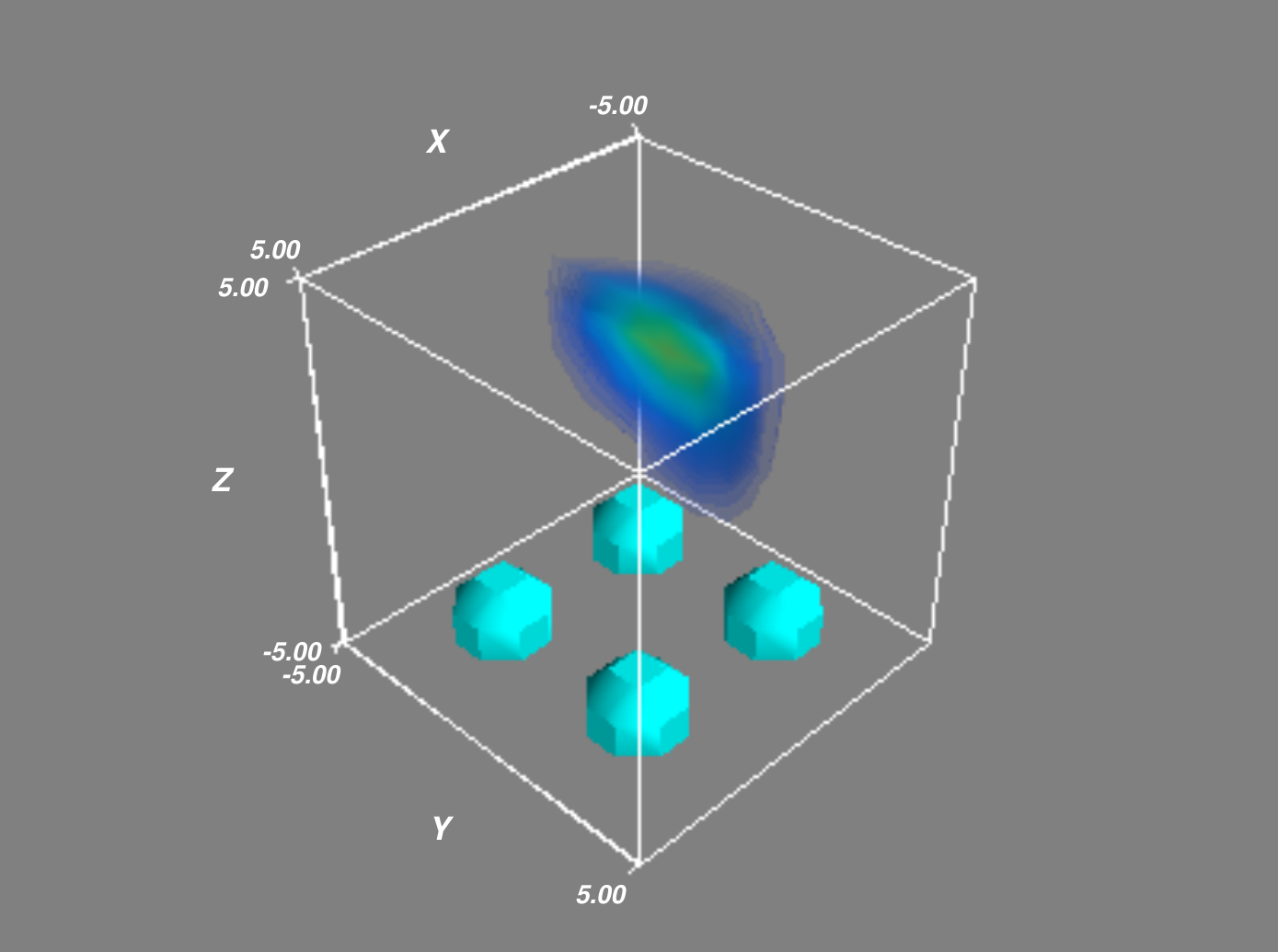}
	\caption{(top) The Boid distribution at a time (time instant $t\simeq 4$) of lower Hellinger distance from the mean-field distribution. (bottom) The mean-field distribution at the same time.}
 \label{fig:boids-4}
\end{figure}

The approximately optimal mean-field distribution does not become nearly as spread as the Boid distribution did. In the beginning, the squared Hellinger distance is very low, as we learned a representation of the initial conditions directly from Boid data. Then, as evolution starts, the squared Hellinger distance attains a peak, as in Fig. \ref{fig:hellinger}. This seems to be due to the spread of the Boid flock as opposed to the mean-field. The most concentrated regions of the Boid flock and the mean-field seem to coincide, as in Fig. \ref{fig:boids-2}. After this initial peak in divergence from each other, the Boid distribution and the mean-field become much more similar as time progresses. It is typical for the two flocks to appear as in Fig. \ref{fig:boids-4}, of course subject to some fluctuations in distance as shown in the plot of the Hellinger distance in Fig. \ref{fig:hellinger}. The times corresponding to Fig. \ref{fig:boids-2} and Fig. \ref{fig:boids-4} are highlighted with the blue and green dots, respectively, in Fig. \ref{fig:hellinger}.
 As mentioned in the previous section, the spatial mesh was taken to be $(11 \times 11 \times 11)$, which was sufficient to recover the dynamics in sufficient detail - flocks with finer features or subflocks obviously require finer grids to capture.

\section{Conclusion}
In this work, we study the inverse problem of identifying complex flocking dynamics in a domain cluttered with obstacles using PDE-constrained optimization methods. 
Our analysis and simulated experiments show that the behavior of cohesive flocks can be recovered accurately with approximate PDE solutions with a relatively coarse mesh in our finite volume method. 
In addition to this result, we developed a simulator for Boids in a bounded region composed of a cube with obstacles formed from cubes, and developed a solver for a variety of Euler alignment system in the same regions.
We performed the optimization in our PDE-constrained optimization problem with a Newton-conjugate gradient method. 
%
%

In current and near future work
we aim to prove that the cost functionals we employed in our learning are differentiable with respect to the parameters providing theoretical guarantees for the convergence of our methods.
We will be addressing the identification and learning problems of interaction (or coordination) laws in MA systems from observed data, albeit from local observations (i.e. with only partial information of the swarm density), and also from a few collaborating observers (sensors). 
In addition we are pursuing similar work for social networks over the Internet and collaborating human-machine swarms.


\bibliographystyle{ieeetran}
\bibliography{references.bib} %

\end{document}